\documentclass[conference,10pt,letterpaper]{IEEEtran}
\usepackage[T1]{fontenc}
\usepackage[utf8]{inputenc}
\usepackage{amsmath,amssymb}
\usepackage{graphicx}
\usepackage{stfloats}
\usepackage{booktabs}
\usepackage{placeins}
\usepackage{microtype}
\usepackage{cite}
\usepackage[hidelinks]{hyperref}

\newcommand{\papertitle}{Beyond Blossom: Accelerating MWPM Decoding for Surface Codes With Local Predecoding and Architecture-Aware Optimizations}
\newcommand{\paperkeywords}{Design automation, algorithms, performance evaluation}

  \newcommand{\paperauthors}{Jiayu Fu, Zirong Shen, Lin Gan, Yiming Zhang, Wenjing Han, Xiaoming Sun, Riling Li, and Guangwen Yang}

\IEEEoverridecommandlockouts

\author{%
  \IEEEauthorblockN{Jiayu Fu\textsuperscript{1,*},
    Zirong Shen\textsuperscript{1},
    Lin Gan\textsuperscript{2},
    Yiming Zhang\textsuperscript{3}\\
    Wenjing Han\textsuperscript{4},
    Xiaoming Sun\textsuperscript{5},
    Riling Li\textsuperscript{5,*},
    Guangwen Yang\textsuperscript{1,4,*}}
  \IEEEauthorblockA{\textsuperscript{1}Department of Computer Science and Technology, Tsinghua University, Beijing, China}
  \IEEEauthorblockA{\textsuperscript{2}Yau Mathematical Sciences Center, Tsinghua University, Beijing, China}
  \IEEEauthorblockA{\textsuperscript{3}School of Physical Sciences, University of Science and Technology of China, Hefei, China}
  \IEEEauthorblockA{\textsuperscript{4}Beijing Academy of Quantum Information Sciences, Beijing, China}
  \IEEEauthorblockA{\textsuperscript{5}Institute of Computing Technology, Chinese Academy of Sciences, Beijing, China}
  \thanks{\textsuperscript{*}Corresponding authors:\protect\newline
    Jiayu Fu (\texttt{fujy24@mails.tsinghua.edu.cn}),\protect\newline
    Riling Li (\texttt{haoliri0@gmail.com}, \texttt{liriling@ict.ac.cn}),\protect\newline
    and Guangwen Yang (\texttt{ygw@mail.tsinghua.edu.cn}).}
}

  \hypersetup{pdftitle={\papertitle},pdfauthor={\paperauthors}}

\usepackage[table]{xcolor}
\usepackage[normalem]{ulem}

\ifdefined\PerfValue\expandafter \fi

\newcommand{\DefinePerfValue}[2]{%
  \expandafter\newcommand\csname BBPerfValue@#1\endcsname{#2}%
}
\DeclareRobustCommand{\PerfValue}[1]{%
  \ifcsname BBPerfValue@#1\endcsname%
    \csname BBPerfValue@#1\endcsname%
  \else%
    \PackageError{performance-macros}{Undefined performance key '#1'}%
      {Declare the key using \string\DefinePerfValue\space before use.}%
  \fi%
}

\DefinePerfValue{legacy.single-core.vs-pymatching.geomean-speedup}{4.23}
\DefinePerfValue{legacy.single-core.vs-pymatching.max-speedup}{5.58}
\DefinePerfValue{legacy.eight-core.vs-pymatching.max-speedup}{10.26}
\DefinePerfValue{legacy.d33.p0p1.beyond-blossom.mean-block-time-us}{21.40}
\DefinePerfValue{legacy.d33.p0p1.beyond-blossom.mean-round-time-us}{0.65}

\DefinePerfValue{xeon.decoder-comparison.p0p05.best.vs-pymatching.max-speedup}{9.38}
\DefinePerfValue{xeon.decoder-comparison.p0p1.best.vs-pymatching.max-speedup}{10.26}
\DefinePerfValue{xeon.decoder-comparison.p0p3.best.vs-pymatching.max-speedup}{7.33}

\DefinePerfValue{xeon.ablation.d33.p0p3.pymatching.mean-time-us}{954.41}
\DefinePerfValue{xeon.ablation.d33.p0p3.structured-simd.mean-time-us}{684.38}
\DefinePerfValue{xeon.ablation.d33.p0p3.structured-simd.vs-pymatching.speedup}{1.39}
\DefinePerfValue{xeon.ablation.d9.p0p05.pymatching.mean-time-us}{1.202}
\DefinePerfValue{xeon.ablation.d9.p0p05.structured-simd.mean-time-us}{1.286}
\DefinePerfValue{xeon.ablation.predecoding.vs-structured-simd.min-speedup}{2.49}
\DefinePerfValue{xeon.ablation.predecoding.vs-structured-simd.max-speedup}{5.19}
\DefinePerfValue{xeon.ablation.d13.p0p1.predecoding.mean-time-us}{1.831}
\DefinePerfValue{xeon.ablation.d13.p0p1.component-parallel.mean-time-us}{2.798}
\DefinePerfValue{xeon.ablation.d33.p0p1.predecoding.mean-time-us}{46.568}
\DefinePerfValue{xeon.ablation.d33.p0p1.component-parallel.mean-time-us}{32.371}
\DefinePerfValue{xeon.ablation.d33.p0p1.component-parallel.vs-predecoding.speedup}{1.44}
\DefinePerfValue{xeon.ablation.d33.p0p3.tail-reset.mean-time-us}{156.040}
\DefinePerfValue{xeon.ablation.d33.p0p3.parallel-predecoding.mean-time-us}{130.151}
\DefinePerfValue{xeon.ablation.d33.p0p3.parallel-predecoding.vs-pymatching.speedup}{7.33}
\DefinePerfValue{xeon.ablation.tail-reset.num-improved-settings}{17}
\DefinePerfValue{xeon.ablation.d33.p0p3.pymatching.L3miss}{0.094}
\DefinePerfValue{xeon.ablation.d33.p0p3.structured-simd.L3miss}{0.017}
\DefinePerfValue{xeon.ablation.d33.p0p3.pymatching.L2miss}{60.212}
\DefinePerfValue{xeon.ablation.d33.p0p3.structured-simd.L2miss}{26.988}

\DefinePerfValue{m4pro.decoder-comparison.p0p05.best.vs-pymatching.max-speedup}{6.36}
\DefinePerfValue{m4pro.decoder-comparison.p0p1.best.vs-pymatching.max-speedup}{6.41}
\DefinePerfValue{m4pro.decoder-comparison.p0p3.best.vs-pymatching.max-speedup}{5.02}
\DefinePerfValue{m4pro.realtime.beyond-blossom.num-supported-settings}{18}
\DefinePerfValue{m4pro.realtime.pymatching.num-supported-settings}{11}

\DefinePerfValue{xeon.predecoder-comparison.d9.p0p3.none.mean-time-us}{8.31}
\DefinePerfValue{xeon.predecoder-comparison.d9.p0p3.none.mean-remaining-syndromes}{34.93}
\DefinePerfValue{xeon.predecoder-comparison.d9.p0p3.coordinate.mean-time-us}{4.72}
\DefinePerfValue{xeon.predecoder-comparison.d9.p0p3.coordinate.mean-remaining-syndromes}{2.52}
\DefinePerfValue{xeon.predecoder-comparison.d9.p0p3.coordinate.matched-percent}{92.79}
\DefinePerfValue{xeon.predecoder-comparison.d9.p0p3.degree-ordered.mean-time-us}{4.00}
\DefinePerfValue{xeon.predecoder-comparison.d9.p0p3.degree-ordered.mean-remaining-syndromes}{1.68}
\DefinePerfValue{xeon.predecoder-comparison.d9.p0p3.degree-ordered.matched-percent}{95.18}
\DefinePerfValue{xeon.predecoder-comparison.d13.p0p3.none.mean-time-us}{18.79}
\DefinePerfValue{xeon.predecoder-comparison.d13.p0p3.none.mean-remaining-syndromes}{110.63}
\DefinePerfValue{xeon.predecoder-comparison.d13.p0p3.coordinate.mean-time-us}{11.40}
\DefinePerfValue{xeon.predecoder-comparison.d13.p0p3.coordinate.mean-remaining-syndromes}{8.44}
\DefinePerfValue{xeon.predecoder-comparison.d13.p0p3.coordinate.matched-percent}{92.37}
\DefinePerfValue{xeon.predecoder-comparison.d13.p0p3.degree-ordered.mean-time-us}{9.10}
\DefinePerfValue{xeon.predecoder-comparison.d13.p0p3.degree-ordered.mean-remaining-syndromes}{5.43}
\DefinePerfValue{xeon.predecoder-comparison.d13.p0p3.degree-ordered.matched-percent}{95.09}
\DefinePerfValue{xeon.predecoder-comparison.d17.p0p3.none.mean-time-us}{41.94}
\DefinePerfValue{xeon.predecoder-comparison.d17.p0p3.none.mean-remaining-syndromes}{253.73}
\DefinePerfValue{xeon.predecoder-comparison.d17.p0p3.coordinate.mean-time-us}{25.19}
\DefinePerfValue{xeon.predecoder-comparison.d17.p0p3.coordinate.mean-remaining-syndromes}{20.04}
\DefinePerfValue{xeon.predecoder-comparison.d17.p0p3.coordinate.matched-percent}{92.10}
\DefinePerfValue{xeon.predecoder-comparison.d17.p0p3.degree-ordered.mean-time-us}{18.80}
\DefinePerfValue{xeon.predecoder-comparison.d17.p0p3.degree-ordered.mean-remaining-syndromes}{12.69}
\DefinePerfValue{xeon.predecoder-comparison.d17.p0p3.degree-ordered.matched-percent}{95.00}
\DefinePerfValue{xeon.predecoder-comparison.d21.p0p3.none.mean-time-us}{81.37}
\DefinePerfValue{xeon.predecoder-comparison.d21.p0p3.none.mean-remaining-syndromes}{485.72}
\DefinePerfValue{xeon.predecoder-comparison.d21.p0p3.coordinate.mean-time-us}{48.93}
\DefinePerfValue{xeon.predecoder-comparison.d21.p0p3.coordinate.mean-remaining-syndromes}{39.23}
\DefinePerfValue{xeon.predecoder-comparison.d21.p0p3.coordinate.matched-percent}{91.92}
\DefinePerfValue{xeon.predecoder-comparison.d21.p0p3.degree-ordered.mean-time-us}{35.72}
\DefinePerfValue{xeon.predecoder-comparison.d21.p0p3.degree-ordered.mean-remaining-syndromes}{24.64}
\DefinePerfValue{xeon.predecoder-comparison.d21.p0p3.degree-ordered.matched-percent}{94.93}
\DefinePerfValue{xeon.predecoder-comparison.d25.p0p3.none.mean-time-us}{143.46}
\DefinePerfValue{xeon.predecoder-comparison.d25.p0p3.none.mean-remaining-syndromes}{828.04}
\DefinePerfValue{xeon.predecoder-comparison.d25.p0p3.coordinate.mean-time-us}{86.65}
\DefinePerfValue{xeon.predecoder-comparison.d25.p0p3.coordinate.mean-remaining-syndromes}{67.93}
\DefinePerfValue{xeon.predecoder-comparison.d25.p0p3.coordinate.matched-percent}{91.80}
\DefinePerfValue{xeon.predecoder-comparison.d25.p0p3.degree-ordered.mean-time-us}{62.43}
\DefinePerfValue{xeon.predecoder-comparison.d25.p0p3.degree-ordered.mean-remaining-syndromes}{42.45}
\DefinePerfValue{xeon.predecoder-comparison.d25.p0p3.degree-ordered.matched-percent}{94.87}
\DefinePerfValue{xeon.predecoder-comparison.d29.p0p3.none.mean-time-us}{236.33}
\DefinePerfValue{xeon.predecoder-comparison.d29.p0p3.none.mean-remaining-syndromes}{1302.48}
\DefinePerfValue{xeon.predecoder-comparison.d29.p0p3.coordinate.mean-time-us}{141.97}
\DefinePerfValue{xeon.predecoder-comparison.d29.p0p3.coordinate.mean-remaining-syndromes}{108.12}
\DefinePerfValue{xeon.predecoder-comparison.d29.p0p3.coordinate.matched-percent}{91.70}
\DefinePerfValue{xeon.predecoder-comparison.d29.p0p3.degree-ordered.mean-time-us}{101.11}
\DefinePerfValue{xeon.predecoder-comparison.d29.p0p3.degree-ordered.mean-remaining-syndromes}{67.37}
\DefinePerfValue{xeon.predecoder-comparison.d29.p0p3.degree-ordered.matched-percent}{94.83}
\DefinePerfValue{xeon.predecoder-comparison.d33.p0p3.none.mean-time-us}{382.49}
\DefinePerfValue{xeon.predecoder-comparison.d33.p0p3.none.mean-remaining-syndromes}{1929.86}
\DefinePerfValue{xeon.predecoder-comparison.d33.p0p3.coordinate.mean-time-us}{223.65}
\DefinePerfValue{xeon.predecoder-comparison.d33.p0p3.coordinate.mean-remaining-syndromes}{161.61}
\DefinePerfValue{xeon.predecoder-comparison.d33.p0p3.coordinate.matched-percent}{91.63}
\DefinePerfValue{xeon.predecoder-comparison.d33.p0p3.degree-ordered.mean-time-us}{156.51}
\DefinePerfValue{xeon.predecoder-comparison.d33.p0p3.degree-ordered.mean-remaining-syndromes}{100.45}
\DefinePerfValue{xeon.predecoder-comparison.d33.p0p3.degree-ordered.matched-percent}{94.79}
\DefinePerfValue{xeon.predecoder-comparison.p0p3.degree-ordered.matched-percent.min}{94.79}
\DefinePerfValue{xeon.predecoder-comparison.p0p3.degree-ordered.matched-percent.max}{95.18}

\title{\papertitle}

\begin{document}
\maketitle

\begin{abstract}
\ifdefined\PerfValue\else
  {}{%
      }}
\fi
Minimum-weight perfect matching (MWPM) is widely used to decode surface codes, but its computational cost increases with code distance. We present Beyond Blossom, a state-of-the-art CPU-based MWPM decoder for surface codes. It combines algorithmic optimization through local predecoding, architecture-aware optimization, and pipeline-wide parallelism to reduce matching workload and execution overhead while preserving MWPM optimality. Across rotated surface codes with distances 9–33 and circuit-level physical error rates of 0.05\%–0.3\%, on a single CPU core, Beyond Blossom achieves a \PerfValue{legacy.single-core.vs-pymatching.geomean-speedup}× geometric-mean speedup and up to \PerfValue{legacy.single-core.vs-pymatching.max-speedup}× speedup over Sparse Blossom in PyMatching. At distance 33 and a physical error rate of 0.1\%, with 8 CPU cores, decoding a block of 33 measurement rounds takes \PerfValue{legacy.d33.p0p1.beyond-blossom.mean-block-time-us} $\mu$s on average, equivalent to \PerfValue{legacy.d33.p0p1.beyond-blossom.mean-round-time-us} $\mu$s per round.

\end{abstract}


\ifdefined\PerfValue\else
  {}{%
      }}
\fi
\section{Introduction}
\label{sec:introduction}

Quantum error correction (QEC) combines QEC codes with decoders~\cite{shor1995scheme,steane1996error,terhal2015quantum}. QEC codes encode logical qubits into multiple physical qubits. A decoder is a classical algorithm that processes repeated stabilizer measurement outcomes, then determines an appropriate recovery operation. The surface code, a QEC code, is well suited to superconducting processors because its stabilizers can be measured using local interactions on a two-dimensional qubit array~\cite{fowler2012surface}. Recent superconducting experiments have demonstrated surface-code memories below threshold, where increasing the code distance reduces the logical error rate~\cite{google2025belowthreshold}. An eight-qubit stability experiment has also demonstrated FPGA decoding with feedback on superconducting hardware~\cite{caune2026realtime}.

These advances place growing demands on classical decoding. Superconducting devices can generate a new round of measurement outcomes about every microsecond~\cite{google2025belowthreshold}.  The decoder must therefore maintain an amortized decoding time below \(1\,\mu\mathrm{s}\) per measurement round to avoid a growing backlog. Logical operations that depend on the decoded result also require low feedback latency~\cite{higgott2025sparse,wu2025micro,caune2026realtime}. Efficient decoding is likewise important in simulations of fault-tolerant quantum circuits, where estimating low logical error rates through Monte Carlo sampling can require decoding millions of circuit samples.

In response, researchers have devoted considerable effort to developing accurate, low-latency surface-code decoders. A surface-code decoder takes a set of detection events as input. A detection event is a change in the measurement outcome of a stabilizer between consecutive rounds. 
Union-find decoding grows and merges clusters of detection events and offers an almost-linear worst-case runtime, but does not generally minimize the total correction weight~\cite{delfosse2021almostlinear}. Tensor-network methods approximate maximum-likelihood decoding by accounting for collections of equivalent errors, with higher accuracy generally requiring more computation~\cite{bravyi2014efficient}. Neural decoders can learn features of experimental noise that are difficult to capture in a simple model, but require suitable training data and sufficiently fast inference~\cite{bausch2024learning}.

Minimum-weight perfect matching (MWPM) casts surface-code decoding as a graph matching problem and is widely used owing to its well-defined optimization objective and mature, efficient implementations~\cite{higgott2025sparse,higgott2022pymatching,wu2023fusion,wu2025micro}.
 Among these implementations, Sparse Blossom~\cite{higgott2025sparse}, implemented in PyMatching 2~\cite{higgott2022pymatching}, performs the matching search directly on the sparse decoding graph.
It avoids the all-to-all shortest-path searches used by some earlier
implementations. Mikami and Yamasaki provide a rigorous theoretical framework for parallel sparse-blossom decoding, but their parallel implementation is slower in practice than single-threaded Sparse Blossom~\cite{mikami2026sparse}.
Fusion Blossom divides the decoding problem into smaller parts, solves them in parallel, and fuses the results into a global minimum-weight matching~\cite{wu2023fusion}.
On FPGAs, Micro Blossom parallelizes dual updates while retaining the primal solver on the CPU. It handles isolated local conflicts while preserving exact MWPM~\cite{wu2025micro}.

These advances suggest two directions for further CPU optimization.
First, at low physical error rates, many detection events occur in adjacent pairs or next to a boundary, allowing local predecoding to reduce the workload passed to the full MWPM solver.
However, heuristic prematching can increase the logical error rate~\cite{smith2023local,alavisamani2024promatch}.
We therefore seek to exploit this locality while preserving MWPM optimality.
Second, further gains may come from more efficient graph data access, single-instruction, multiple-data (SIMD) execution~\cite{flynn1972organizations}, and reduced overhead for thread coordination and state reinitialization.

Building on these investigations, we present \textbf{Beyond Blossom}\footnote{Beyond Blossom, https://github.com/Edgarrrrrr/beyond-blossom}, an exact MWPM decoder that combines algorithmic optimization through predecoding, architecture-aware optimization, and pipeline-wide parallelism on modern CPUs. Together, these techniques accelerate both the blossom search and the surrounding decoding stages while preserving MWPM optimality. We provide both single-core and multi-core implementations of Beyond Blossom. Across rotated surface-code memory circuits with odd distances from 9 to 33 and circuit-level physical error rates of 0.05\%, 0.1\%, and 0.3\%, the single-core configuration achieves a \textbf{$\PerfValue{legacy.single-core.vs-pymatching.geomean-speedup}\times$ geometric-mean speedup} and up to \textbf{$\PerfValue{legacy.single-core.vs-pymatching.max-speedup}\times$ speedup} over unmodified PyMatching 2.3.1\footnote{PyMatching is not limited to surface-code decoding, but our performance comparisons in this paper focus exclusively on surface codes.}, a prior state-of-the-art CPU MWPM decoder that supports only single-threaded decoding~\cite{higgott2022pymatching,higgott2025sparse}. Using eight physical CPU cores, Beyond Blossom achieves up to a \textbf{$\PerfValue{legacy.eight-core.vs-pymatching.max-speedup}\times$ decoding speedup} over single-threaded PyMatching.
Our main contributions are:

\begin{enumerate}
\item \textbf{Degree-ordered one-hop predecoding.}
We introduce a predecoder that reduces global matching work while preserving MWPM optimality.
It processes common local error patterns in degree order while maintaining dual feasibility, then transfers the resulting matching and dual state directly to the full solver.

\item \textbf{SIMD-enabled structured grid representation.}
We exploit the regular local connectivity of surface-code decoding graphs to introduce a structured grid representation. Its data layout improves cache locality, reduces pointer chasing and irregular memory accesses, and enables SIMD execution of key neighborhood computations in the dual phase.

\item \textbf{Pipeline-wide CPU parallelism for short blocks.}
We develop a parallel design for decoding blocks of \(d\) measurement rounds in a distance-\(d\) surface code. It combines parallel predecoding, connected-component parallelism, and overlapped state reset across thread groups.
\end{enumerate}

\section{Preliminaries}
\label{sec:preliminary}

This section briefly introduces the background and notation needed to describe Beyond Blossom. We first review surface-code QEC, then introduce decoding graphs and the minimum-weight perfect matching (MWPM) formulation. Finally, we outline the alternating-tree operations and primal--dual framework of blossom-based decoders.

\subsection{Surface-Code Quantum Error Correction}

We consider the rotated surface code, which encodes one logical qubit using \(d^2\) data qubits and $d^2-1$ ancilla qubits~\cite{horsman2012latticesurgery}. Here, the code distance \(d\) is the minimum weight of a nontrivial logical Pauli operator. With ideal stabilizer measurements, a distance-\(d\) code can correct any Pauli error affecting at most \(\lfloor\frac{d-1}{2}\rfloor\) data qubits~\cite{dennis2002memory}. 

\begin{figure}[!htbp]
    \centering
    \includegraphics[width=\linewidth]{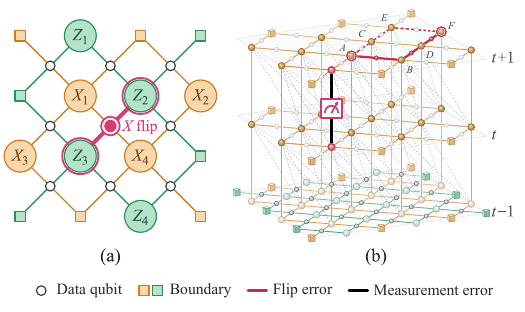}
    \caption{Rotated surface-code layout and 3D decoding graph. (a) A distance-3 code with 9 data qubits (white circles) and 8 ancilla qubits for measuring \(X\)-type (orange) and \(Z\)-type (green) stabilizers. (b) Three middle layers of the decoding graph for a distance-5 surface code, emphasizing the \(X\)-stabilizer subgraph; the \(Z\)-stabilizer subgraph is constructed analogously. }
    \label{fig:surface-code}
\end{figure}

The surface code detects errors through repeated measurements of local \(X\)- and \(Z\)-type stabilizers on a two-dimensional lattice~\cite{dennis2002memory,kitaev2003fault,fowler2012surface}.
In the rotated layout illustrated in Fig.~\ref{fig:surface-code}(a), each stabilizer is a product of Pauli \(X\) or \(Z\) operators acting on four neighboring data qubits in the interior or two at a boundary. Each stabilizer is measured using an ancilla qubit, which interacts with the corresponding data qubits through entangling gates and is then read out to yield a \(\pm1\) outcome.
An odd number of \(X\) errors on the data qubits of a \(Z\)-type stabilizer flips its measurement outcome, and \(Z\) errors affect \(X\)-type stabilizers in the same way. In the illustrated example, an \(X\) error on the highlighted data qubit flips the measurement outcomes obtained from the adjacent \(Z_2\) and \(Z_3\) stabilizer checks.

The logical error rate per measurement cycle, \(\varepsilon_d\), depends on both the code distance \(d\) and the physical error rate \(p\). For odd \(d\), it approximately scales below threshold as~\cite{fowler2012surface}
\[
    \varepsilon_d \propto
    \left(\frac{p}{p_{\mathrm{thr}}}\right)^{(d+1)/2},
\]
where \(p_{\mathrm{thr}}\) is the threshold for the chosen noise model and decoder. At a fixed \(p<p_{\mathrm{thr}}\), this scaling indicates that increasing \(d\) reduces \(\varepsilon_d\) exponentially, a trend also demonstrated experimentally~\cite{google2025belowthreshold}.

Representative superconducting experiments have demonstrated surface-code memories at distances \(d=3,5,7\)~\cite{google2025belowthreshold}. Practical quantum computations, however, can require larger code distances to keep the overall failure probability low over many logical operations. For example, resource estimates for quantum dynamics simulation, chemical energy estimation, and RSA-2048 factorization demand code distances ranging from \(d=9\) to \(33\) at assumed physical error rates of \(p=10^{-4}\) and \(10^{-3}\)~\cite{beverland2022assessing}. These larger code distances increase the volume of stabilizer measurement data and hence the decoding workload, motivating decoder acceleration for practical fault-tolerant quantum computation.

\subsection{Decoding Graph}

Decoding aims to determine an appropriate recovery operation from the observed stabilizer measurement outcomes. In addition to the data-qubit errors described above, stabilizer measurements can themselves be faulty. We therefore decode measurement histories spanning \(d\) rounds to correct up to \(\lfloor(d-1)/2\rfloor\) data and measurement faults~\cite{dennis2002memory}.
Under an independent-error model, we represent surface-code decoding using a three-dimensional decoding graph \(G=(V,E)\), as shown in Fig.~\ref{fig:surface-code}(b).
Each layer corresponds to one measurement round. Each non-boundary vertex is called a detector; the detector at \((x,y,t)\) corresponds to the stabilizer measurement at spatial location \((x,y)\) in round \(t\). Each edge represents a possible error mechanism, connecting either the two detectors it flips or a single detector to a virtual boundary. Within-layer edges represent data errors, while errors arising during measurement are represented by interlayer edges. For example, the black edge in Fig.~\ref{fig:surface-code}(b) represents a measurement error and connects two detectors associated with the same stabilizer.

A detector with a detection event is called a defect. The same set of defects can arise from different error configurations. In the \(t+1\) layer of Fig.~\ref{fig:surface-code}(b), the solid red chain \(A\!-\!B\!-\!D\!-\!F\) flips the outcomes of internal detectors \(B\) and \(D\) twice each, producing defects only at \(A\) and \(F\). These defects alone cannot distinguish between the error chains \(A\!-\!B\!-\!D\!-\!F\) and \(A\!-\!C\!-\!E\!-\!F\), whose probabilities may differ. The objective of MWPM decoding is to identify the most probable set of error edges consistent with the observed defect set.

Assuming independent edge errors, a path \(\mathit{\gamma}\) has
probability
\begin{equation}
    P(\mathit{\gamma})
    =\prod_{e\in \mathit{\gamma}}p_e
     \prod_{e\in E\setminus \mathit{\gamma}}(1-p_e)
    \propto
    \prod_{e\in \mathit{\gamma}}\frac{p_e}{1-p_e}.
\end{equation}
Each edge with error probability \(p_e\) is assigned the weight
\(w_e=\ln[(1-p_e)/p_e]\), making probability maximization equivalent
to minimizing the total edge weight.

We reduce decoding to MWPM by constructing an auxiliary matching graph \(G_M=(V_M,E_M)\)~\cite{dennis2002memory,wang2003confinement}. Here, \(V_M\) is the set of all defects, and \(E_M\) connects every pair of defects \(u,v\). Each matching edge is weighted by their shortest-path distance in the decoding graph \(G\):
\begin{equation}
    w^M_{uv}=\operatorname{dist}_G(u,v).
\end{equation}
We then add virtual boundary vertices and edges so that defects can independently match to a boundary. MWPM pairs every vertex exactly once with minimum total edge weight, yielding a most probable error configuration consistent with the observed defect set. The matched pairs are then expanded into their corresponding paths in the decoding graph. Finally, these paths are combined modulo two to obtain the inferred error set.
\subsection{Blossom-Based Decoders}

\begin{figure}[!htbp]
    \centering
\includegraphics[width=1\linewidth]{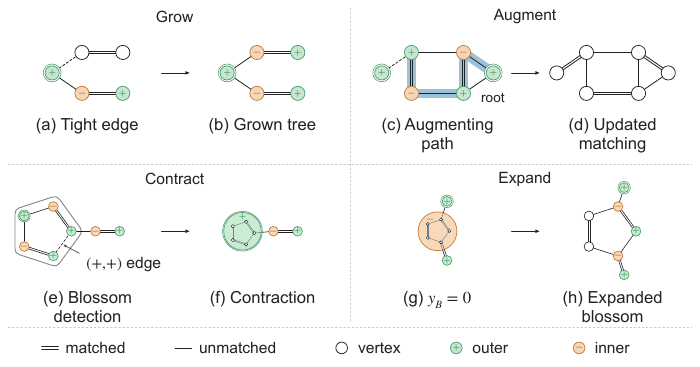}
    \caption{Key steps of the blossom algorithm for MWPM. Dashed edges indicate edges that became tight during the immediately preceding dual update. (a)–(b) Alternating-tree growth. (c)–(d) Augmentation yields one additional matched pair. (e)–(f) Blossom contraction into a pseudo-node. (g)–(h) Blossom expansion.}
    \label{fig:blossom}
\end{figure}

Edmonds’ Blossom algorithm provides a polynomial-time solution to MWPM in general graphs~\cite{edmonds1965paths,edmonds1965polyhedron,kolmogorov2009blossomv}. The algorithm alternates between primal and dual updates. Primal updates modify the matching and alternating trees, while dual updates enable further search without violating the linear constraints. The blossom algorithm is guided by a primal–dual formulation on the graph \(G_M=(V_M,E_M)\). For each edge \(e\in E_M\), let \(x_e\) denote its selection variable. Let \(\delta(B)\) denote the edges in \(E_M\) crossing between \(B\) and \(V_M\setminus B\), with \(\delta(v)=\delta(\{v\})\), and \(\mathcal O\) contain the odd-sized subsets of \(V_M\) with at least three vertices. The primal linear program is

\begin{equation}
\begin{aligned}
\min_x\quad
&\sum_{e\in E_M}w_ex_e\\
\text{s.t.}\quad
&\sum_{e\in\delta(v)}x_e=1,
&&\forall v\in V_M,\\
&\sum_{e\in\delta(B)}x_e\geq1,
&&\forall B\in\mathcal O,\\
&x_e\geq0,
&&\forall e\in E_M.
\end{aligned}
\end{equation}
The corresponding dual assigns unrestricted variables \(y_v\) to vertices and nonnegative variables \(y_B\) to odd sets:
\begin{equation}
\begin{gathered}
\max_y\quad
\sum_{v\in V_M}y_v+\sum_{B\in\mathcal O}y_B\\
\text{s.t.}\quad
\operatorname{slack}(e)\geq0,\quad \forall e\in E_M,\\
y_B\geq0,\quad \forall B\in\mathcal O.
\end{gathered}
\end{equation}
For an edge \(e=(u,v)\), its slack is defined as
\begin{equation}
\operatorname{slack}(e)\equiv w_e-y_u-y_v
-\sum_{\substack{B\in\mathcal O\\e\in\delta(B)}}y_B.
\end{equation}
We call an edge \textbf{tight} when \(\operatorname{slack}(e)=0\).

The algorithm starts with every vertex as an unmatched root labeled \(+\). During execution, the algorithm maintains an alternating forest rooted at unmatched nodes. Paths from each root alternate between unmatched and matched edges. Nodes at even and odd depths are labeled \(+\) (outer) and \(-\) (inner), respectively. Each node represents either an original vertex or a contracted blossom. Nodes outside the alternating forest are frozen; their dual variables remain unchanged during the dual phase.

With the dual variables fixed, the primal search uses only tight edges and performs four operations: GROW, AUGMENT, SHRINK, and EXPAND. GROW follows a tight edge from an outer node to an unlabeled matched node outside the forest, adding that node as inner and its matched partner as outer (Fig.~\ref{fig:blossom}(a)–(b)). AUGMENT applies when a tight edge joins outer nodes in different trees, yielding an augmenting path between their unmatched roots. Flipping matched and unmatched edges along this path increases the matching size by one (Fig.~\ref{fig:blossom}(c)–(d)). SHRINK applies when a tight edge joins two outer nodes in the same tree, closing an odd cycle called a blossom. The algorithm contracts this cycle into a pseudo-node with a new dual variable initially set to zero (Fig.~\ref{fig:blossom}(e)–(f)). EXPAND is triggered when the dual variable \(y_B\) of an inner blossom \(B\) reaches zero. The blossom is expanded into its constituent nodes, and the affected alternating-tree path is reconstructed without changing the matching (Fig.~\ref{fig:blossom}(g)–(h)).  The primal phase therefore involves complex updates to matching and alternating-tree structures.

When no further primal operation is available, the algorithm performs a dual update. Under the fixed-\(\delta\) strategy, all active trees take the same step: outer-node duals increase by \(\delta\), inner-node duals decrease by \(\delta\), and variables inside contracted blossoms remain fixed. The step ends at the earliest eligible event across all active trees: an edge becomes tight or an inner blossom’s dual reaches zero. A global priority queue orders candidate events by their scheduled times, finding the earliest event. The primal search then resumes at the triggering edge or blossom. Primal operations may switch nodes among \(+\), \(-\), and frozen states or modify the blossom structure. Any new valid event candidates produced by these changes are added to the priority queue~\cite{kolmogorov2009blossomv}. The algorithm terminates when the priority queue is empty, at which point the complementary slackness conditions guarantee that the resulting matching is an MWPM~\cite{kolmogorov2009blossomv}.

Although only defects require pairing in QEC decoding, the Sparse Blossom algorithm searches directly on the decoding graph \(G\) using graph-fill regions (hereafter referred to as regions)~\cite{higgott2025sparse}.
Each defect \(u\) initially seeds a region that grows along weighted edges under the control of its radius \(y_u\), potentially reaching nondefect detectors. The radius \(y_u\) is the dual variable associated with defect \(u\). When two regions collide, the relevant dual contributions equal the weight of the corresponding pairing edge, making it tight. At a region’s frontier, growth adds ordinary detectors and shrinkage releases them. These ownership changes and region collisions are handled as events in a global priority queue. This implements the Blossom search through region growth, shrinkage, and collision detection, without first computing all pairwise shortest-path distances or constructing the complete matching graph \(G_M\).

\section{Proposed Method}
\label{sec:method}

MWPM decoding performance depends on both the matching workload and the efficiency of its CPU implementation.
At low physical error rates, sparse local errors and the regular structure of surface-code graphs offer opportunities to improve both.
Beyond Blossom therefore combines algorithmic optimization through predecoding (Section~\ref{sec:predecoding}), architecture-aware optimization (Section~\ref{sec:structured-grid-simd}), and pipeline-wide parallelism (Section~\ref{sec:multilevel-parallelism}) to reduce decoding  time while preserving MWPM optimality.

\subsection{Degree-Ordered One-Hop Predecoding}
\label{sec:predecoding}

At low physical error rates, many common errors produce adjacent defect pairs or defects that can be matched directly to a boundary~\cite{smith2023local,wu2025micro}.
These patterns occupy small local neighborhoods and often admit simple matching relationships.
Under an independent local fault model, two distinct faults occur within a fixed-size neighborhood with probability \(O(p^2)\), providing substantial opportunities for local processing.

\begin{figure*}[!tbp]
  \centering
  \includegraphics[width=\textwidth]{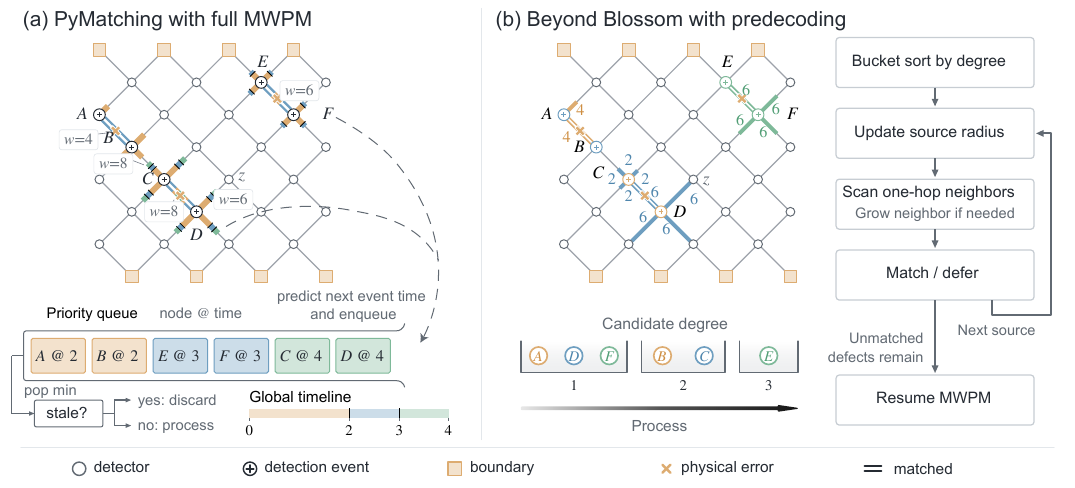}
  \caption{
    Event-driven MWPM decoding in Sparse Blossom and degree-ordered one-hop predecoding in Beyond Blossom.
    (a) Sparse Blossom schedules region events through a shared priority queue and discards stale events when dequeued.
    (b) Beyond Blossom's one-hop predecoder processes defects in increasing order of candidate degree, bounds radius growth using the slack of all incident edges, and accepts tight-edge matches.
    If all defects are matched, the optimal solution is returned immediately.
    Otherwise, the partial matching and dual state initialize the full MWPM solver.}
  \label{fig:predecode}
\end{figure*}

Nevertheless, a full MWPM solver maintains general-purpose search and scheduling structures even for these simple patterns.
Sparse Blossom~\cite{higgott2025sparse}, for example, advances growing regions uniformly along a shared timeline and processes events through a global priority queue.
Events from spatially distant regions therefore share the same ordering mechanism.
State changes can also invalidate queued events: after \(A\) and \(B\) match at time 2 in Fig.~\ref{fig:predecode}(a), a previously scheduled growth event for \(B\) must be revalidated when dequeued.
Alternating-tree maintenance, event-time calculations, queue operations, and validity checks collectively increase the cost of resolving local pairs.

We therefore introduce a degree-ordered one-hop predecoder that quickly matches most defects through simple local operations, constructing a matching and a feasible dual state before invoking the MWPM solver when needed.
Figure~\ref{fig:predecode}(b) illustrates the algorithm.

The one-hop restriction defines the scope of local processing.
\begin{enumerate}
  \item \textbf{Direct matching.}
  We match an unmatched defect to another unmatched defect or a boundary along a tight edge, then freeze the matched defects' radii.
 
  \item \textbf{Growth limited by a neighbor.}
  When an edge to a nondefect detector or a matched defect becomes tight, the source cannot grow further.
\end{enumerate}

These rules yield two structural simplifications:
\begin{enumerate}
  \item \textbf{Flat dual state.}
  Because the predecoder neither extends alternating trees nor contracts blossoms, each defect retains its own dual radius without a blossom hierarchy.
  We therefore store radii in a flat array and check edge tightness directly from the two endpoint radii and the edge weight.

  \item \textbf{Sequential growth without a global event queue.}
  Radii only increase or remain frozen, so the predecoder has no shrinking (\(-\)) nodes.
  In a general blossom solver, an edge \(e=(u,v)\) between a growing (\(+\)) node \(u\) in tree \(T\) and a shrinking node \(v\) in another tree \(T'\) imposes the cross-tree constraint \(\delta_T-\delta_{T'}\leq\operatorname{slack}(e)\), where \(\delta_T\) and \(\delta_{T'}\) denote the dual increments of the two trees~\cite{kolmogorov2009blossomv}.
  This class of constraints is absent from predecoding.
  We can therefore update nodes sequentially by different amounts, bounding each update by the current incident-edge constraints to preserve dual feasibility.
  This permits a direct traversal of the defects without equal-rate advancement, event-time prediction, or a global priority queue.
\end{enumerate}

Degree ordering determines the processing sequence.
Let \(S\subseteq V\) denote the input defect set in the decoding graph \(G\), excluding virtual boundaries.
We define the candidate degree of defect \(u\) as \(k(u)=|N(u)\cap S|+b(u)\),
where \(N(u)\) contains its ordinary neighboring nodes and \(b(u)\) counts its valid boundary candidates.
A single stable bucket sort orders defects by increasing candidate degree.
For each defect \(u\), let \(r_u\geq0\) denote its dual radius, corresponding to \(y_u\) in the absence of blossom contributions.
All radii are initialized to zero.
For each unmatched source \(u\), we update its radius as
\begin{equation}
  r_u \mathrel{\text{}}\leftarrow
  \min_{v:(u,v)\in E}\bigl(w_{uv}-r_v\bigr),
  \label{eq:predecode-cap}
\end{equation}
where \(w_{uv}\) is the weight of edge \((u,v)\) in \(G\). The minimum includes all incident edges.
We take \(r_v=0\) for nondefect nodes and virtual boundaries.
We then examine unmatched defect neighbors.
If \(r_u+r_v=w_{uv}\), the edge is tight and the pair is immediately matched.
Otherwise, we attempt to grow \(v\) using~\eqref{eq:predecode-cap} and accept the pair if the edge becomes tight.
If no defect match is found, we check for a tight boundary edge.
If the source remains unmatched, we retain its radius and proceed to the next source.

These updates preserve \(r_u+r_v\leq w_{uv}\).
Summing these constraints along paths gives a feasible dual solution on \(G_M\), with all blossom dual variables zero.
Once every defect is matched exactly once through tight edges, complementary slackness guarantees optimality~\cite{kolmogorov2009blossomv}, allowing immediate return.
Otherwise, we consistently import the radii and matching relationships into the full MWPM solver.
Only unmatched defects become initial alternating-tree roots and initiate scheduling.
The full MWPM solver can undo and revise prematches as it completes the matching.

The effectiveness of one-hop predecoding on surface codes also follows from the structure of the edge weights.
For a decoding-graph edge \(e\in E\) dominated by first-order faults, its error probability satisfies \(p_e=a_ep+O(p^2)\), where \(a_e>0\) is a coefficient independent of \(p\). Using the edge-weight definition above, as \(p\to0\), \(w_e=\ln[(1-p_e)/p_e]=\ln(1/p)-\ln a_e+O(p)\).
Such edges share the leading term \(\ln(1/p)\), with differences primarily determined by their local coefficients.
Consider an unprocessed adjacent defect pair \(u,v\) with no other neighboring defects.
Let \(\mu_u\) and \(\mu_v\) denote the minimum incident-edge weights at the respective endpoints, including edges to nondefect nodes and boundaries.
If \(u\) is processed first, its radius reaches \(\mu_u\). The pair becomes tight when \(v\) supplies the remaining radius \(w_{uv}-\mu_u\), which is feasible exactly when
\begin{equation}
  \mu_u+\mu_v\geq w_{uv}.
  \label{eq:predecode-pair-condition}
\end{equation}
When the target edge and the neighboring edges limiting both endpoints are dominated by first-order faults, \(\mu_u+\mu_v-w_{uv}=\ln(1/p)+O(1)\).
This condition holds at sufficiently low error rates, allowing sequential radius updates to match the isolated pair.

Degree ordering further exploits the asymmetry of these updates.
When local edge weights are similar, the source processed first typically acquires a large radius, while its partner supplies only a small remainder.
The larger source radius restricts the growth available to its other defect neighbors.
Prioritizing sources with fewer candidates therefore reduces the number of other defects directly affected by each large update.
For example, in Fig.~\ref{fig:predecode}(b), matching \((B,C)\) first leaves \(A\) and \(D\) unmatched during predecoding, whereas processing the degree-one sources \(A\) and \(D\) first yields \((A,B)\) and \((C,D)\).
A single static ordering preserves more opportunities for subsequent matching at low preparation cost.

For the fixed-size neighborhoods of the surface-code graph, bucket sorting and one-hop growth and matching require \(O(|S|)\) total work.
Fully covered inputs terminate immediately, while the remaining inputs enter the full solver with fewer unmatched roots, reducing computational work while preserving MWPM optimality.

\subsection{SIMD-Enabled Structured Grid Representation}
\label{sec:structured-grid-simd}

The blossom algorithm alternates between primal and dual phases, which exhibit distinct computational characteristics.
The primal phase maintains the alternating forest and matching structures through tree growth, augmentation, and blossom contraction and expansion.
These operations involve extensive data-dependent branching and serially dependent pointer accesses.
Their execution efficiency therefore relies heavily on CPU branch prediction and out-of-order execution, while offering limited scope for software transformations based on regular computation and batched parallel execution.
In contrast, the dual phase, which is a major computational hotspot~\cite{wu2025micro}, exhibits a more regular computational structure.
It repeatedly evaluates edge slack to determine admissible dual-variable updates and identify the next edge to become tight.
For a given dual state, candidate edges can be evaluated independently, followed by a minimum reduction to determine the next update.
This repeated local computation provides opportunities for data-layout optimization, instruction-level parallelism, and SIMD vectorization.

\begin{figure}[!htbp]
  \centering
  \includegraphics[width=\columnwidth]{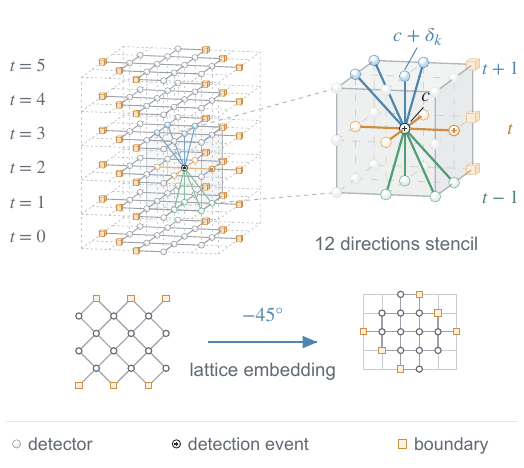}
  \caption{Structured lattice embedding of a surface-code decoding graph.
    A twelve-direction stencil represents local connectivity, allowing
    neighbors of site \(c\) to be addressed as \(c+\delta_k\).}
  \label{fig:structured-grid}
\end{figure}

The local stabilizer checks and repeated measurements of surface codes provide a natural three-dimensional spacetime representation for their decoding graphs.
Under local circuit-level noise, each fault can trigger detection events only within a bounded spacetime neighborhood, allowing graph edges to be described by a finite set of relative displacements.
As illustrated in Fig.~\ref{fig:structured-grid}, we apply an affine transformation to detector coordinates and embed each independently decoded X/Z component of the rotated surface code in a regular three-dimensional lattice.
The spatial, temporal, and diagonal spacetime edges induced by circuit-level noise are then represented by a twelve-direction stencil specific to each component.
This representation encodes adjacency through fixed lattice offsets~\cite{datta2009stencil}.
For lattice coordinates \((x,y,t)\), let \((\Delta x_k,\Delta y_k,\Delta t_k)\) denote the displacement in direction \(k\in\{0,\ldots,11\}\). The linear site index and neighbor offset are
\begin{equation}
  \begin{aligned}
    c &= x+N_x(y+N_y t),\\
    \delta_k &= \Delta x_k+N_x\Delta y_k+N_xN_y\Delta t_k,
  \end{aligned}
  \label{eq:simd-lattice-offsets}
\end{equation}
where \(N_x\) and \(N_y\) denote the lattice extents along the two spatial dimensions.
For an existing edge in direction \(k\), its neighbor is located at index \(c+\delta_k\).
The fixed directions and offsets also permit full unrolling of the neighbor scan, allowing the compiler to interleave address calculations and candidate evaluations across neighbors to exploit instruction-level parallelism.

Building on this addressing scheme, we organize data according to the access requirements of neighbor scanning.
To reduce cache pollution and improve cache-line utilization, we compactly store the edge weights, each occupied detector's top-level region identifier, and radius-related state repeatedly accessed during candidate evaluation.
Information for event maintenance, logical-observable tracking, and path recovery is stored separately.
This hot--cold separation concentrates the scan's accesses on the required state and reduces the unrelated data fetched with each cache line.

The structured layout further enables SIMD vectorization across neighbor candidates.
Each vector lane processes a different neighboring direction of the same source node, sharing the source state while independently evaluating a candidate event time.
However, a scalar scan follows conditional branches based on edge presence, region ownership, and growth state, and different neighbors may satisfy different conditions.
To express these differences within a uniform SIMD instruction stream, we replace per-neighbor conditional control flow with mask operations and organize the scan into arithmetic and eligibility paths.

The arithmetic path evaluates candidate times in parallel.
At the current dual time, let \(r_u\) and \(r_{v_k}\) denote the local effective radii at source site \(u\) and its neighbor \(v_k\), and let \(w_k=w_{uv_k}\) denote their edge weight.
We set \(h_k=1\) when both endpoints are growing and \(h_k=0\) otherwise.
The mask \(m_k\) selects existing edges with a positive combined growth rate, excluding edges internal to the same top-level region. Candidate time increments are
\begin{equation}
  \begin{aligned}
    \widehat{\Delta T}_k
      &= \frac{w_k-r_u-r_{v_k}}{1+h_k},\\
    \Delta T_{\min}
      &= \min_{0\le k<\text{}12}
         \operatorname{select}(m_k,\widehat{\Delta T}_k,+\infty).
  \end{aligned}
  \label{eq:simd-candidate-selection}
\end{equation}
For eligible candidates, the numerator is the remaining slack and the denominator is its closing rate.
Applying the validity mask before reduction replaces invalid candidates with infinity, eliminating per-neighbor conditional branches within each batch.
Although uniform evaluation includes some candidates that scalar branches would discard early, it permits multiple neighbors to be evaluated in parallel.
We use 256-bit AVX-512VL to evaluate four 64-bit candidates per batch, covering all twelve directions in three batches.
The benefit of four-way parallel evaluation outweighs the additional computation on invalid candidates.

\subsection{Pipeline-Wide CPU Parallelism for Short Decoding Blocks}
\label{sec:multilevel-parallelism}

\begin{figure*}[!tbp]
  \centering
  \includegraphics[width=\textwidth]{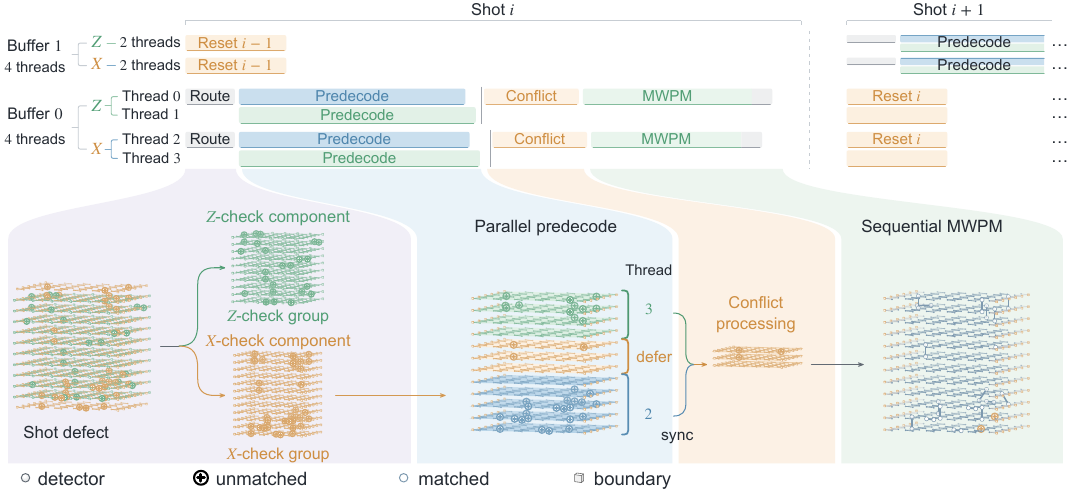}
  \caption{Pipeline-wide CPU parallelism for short decoding blocks.
    Separate thread groups decode independent components concurrently.
    Within each component, two threads predecode disjoint temporal interiors,
    then synchronize before one thread processes the deferred conflict-band
    sources and, if needed, continues with MWPM.
    Two alternating workspaces overlap reset of the previous shot with
    decoding the next.}
  \label{fig:multilevel-parallelism}
\end{figure*}

The central challenge in thread-level MWPM parallelization on the decoding graph is to partition the graph while preserving dual feasibility across partition boundaries during independent execution.
Fusion Blossom~\cite{wu2023fusion} constrains local growth using temporary partition boundaries, then fuses the local matching states and resumes joint solving.
Clustering-based approaches~\cite{mikami2026sparse} instead preprocess the defect set to identify clusters of defects that can be processed independently during the current parallel stage.
For short decoding blocks, limited computation is available to amortize these preparation and coordination costs, which can offset the  time savings from parallel execution.
Predecoding further reduces the remaining MWPM workload.
We therefore retain sequential MWPM within each component and organize pipeline-wide CPU parallelism around independent components, state cleanup, and local predecoding.

First, the independently decoded \(X/Z\) connected components can be solved concurrently before their outputs are combined.
We identify connected components using a disjoint-set data structure during decoder construction and assign each component a separate thread group and event priority queue, allowing independent event streams to advance concurrently.
As decoding time decreases, reset operations---including restoring decoding-graph state and clearing defect and region state from the preceding shot---also become a significant cost.
We overlap this work through double buffering: once the current result has been extracted and copied out, its workspace is reset while the next input is decoded in the other workspace.
Figure~\ref{fig:multilevel-parallelism} illustrates the overall thread organization and execution timeline.

Building on this organization, we further parallelize predecoding within each component.
In our tests, predecoding accounts for a substantial fraction of total decoding time, making it an important target for further acceleration.
Compared with the dynamically evolving MWPM search, predecoding accesses a fixed neighborhood, allowing dependencies between concurrent operations to be determined from the local structure of the decoding graph.

Local operations, however, are not inherently independent.
A predecoding operation can update one-hop neighbors, while candidate evaluation may read state up to two hops away.
Operations with overlapping neighborhoods can therefore introduce read--write conflicts.
Using the structured lattice, we partition each component along the measurement-time dimension and retain a three-layer conflict band around the partition boundary, as illustrated in Fig.~\ref{fig:multilevel-parallelism}.
Each stencil edge spans at most one measurement round.
Deferring operations originating in this band therefore prevents either interior task from modifying state accessed by the other.
Two threads first process their respective interiors concurrently, after which operations originating in the conflict band are executed sequentially.
The matching and radius states produced by the partitions then jointly initialize MWPM on the complete component, which continues the search to complete the remaining matching.

\section{Experimental Evaluation}
\label{sec:evaluation}

We evaluate Beyond Blossom primarily on an Intel Xeon server, examining exact MWPM correctness (Section~\ref{sec:eval-correctness}), amortized decoding time (Section~\ref{sec:eval-comparison}), the contributions of algorithmic and architecture-aware optimizations and parallel execution (Section~\ref{sec:eval-ablation}), and the mechanisms behind the speedup (Section~\ref{sec:eval-predecoding}).
We further evaluate portability to the Arm-based Apple M4 Pro and its ability to meet real-time decoding requirements (Section~\ref{sec:eval-arm}).

\subsection{Experimental Setup}
\label{sec:eval-setup}

We use rotated surface codes with \(d\in\{9,\allowbreak13,\allowbreak17,\allowbreak21,\allowbreak25,\allowbreak29,\allowbreak33\}\), with each decoding block containing \(d\) measurement rounds.
Physical error rates are \(p\in\{0.05\%,\allowbreak0.1\%,\allowbreak0.3\%\}\), yielding 21 settings.
These distances cover larger code sizes relevant to resource estimates for fault-tolerant quantum computation~\cite{beverland2022assessing}, while the error rates span workloads ranging from sparse local errors to higher defect densities.
We generate rotated surface-code memory-X circuits using Stim~\cite{gidney2021stim} under circuit-level noise.
The after-Clifford depolarization, before-round data depolarization, before-measurement flip, and after-reset flip probabilities are all set to \(p\).
Each input contains defect data from both the \(X\)- and \(Z\)-check components.

Our experiments use a dual-socket Intel Xeon Gold 6230R server with 52 physical cores and a 14-core Apple M4 Pro.
Both builds use C++20 and \texttt{-O3}, with GCC 12.2.0 on Xeon and Apple Clang 21.0.0 on M4 Pro.
The Xeon and M4 Pro builds use AVX-512VL/AVX2 and NEON instructions, respectively.

\subsection{Correctness}
\label{sec:eval-correctness}

Beyond Blossom returns exactly the same total matching weight as PyMatching for every tested input.
These results agree with our optimality analysis in Section~\ref{sec:predecoding} and validate Beyond Blossom's exact MWPM decoding.

\subsection{Comparison with Open-Source Decoders}
\label{sec:eval-comparison}

We compare Beyond Blossom with PyMatching 2.3.1~\cite{higgott2025sparse}, Parity Blossom, Fusion Blossom~\cite{wu2023fusion} from fusion-blossom 0.2.13, and two Union-Find implementations, UnionFind~\cite{UnionFindCPP} and the union-find implementation in the ldpc library 2.4.1 (LDPC-UF). PyMatching is invoked through its native C++ interface, avoiding Python call overhead.  All baselines use one physical core except Fusion Blossom, which is restricted to four physical cores. Timing includes per-shot preparation, decoding, output extraction, and state reset. Setup, warm-up, I/O, and validation are excluded.

\begin{figure}[!htbp]
  \centering
  \includegraphics[width=\linewidth]{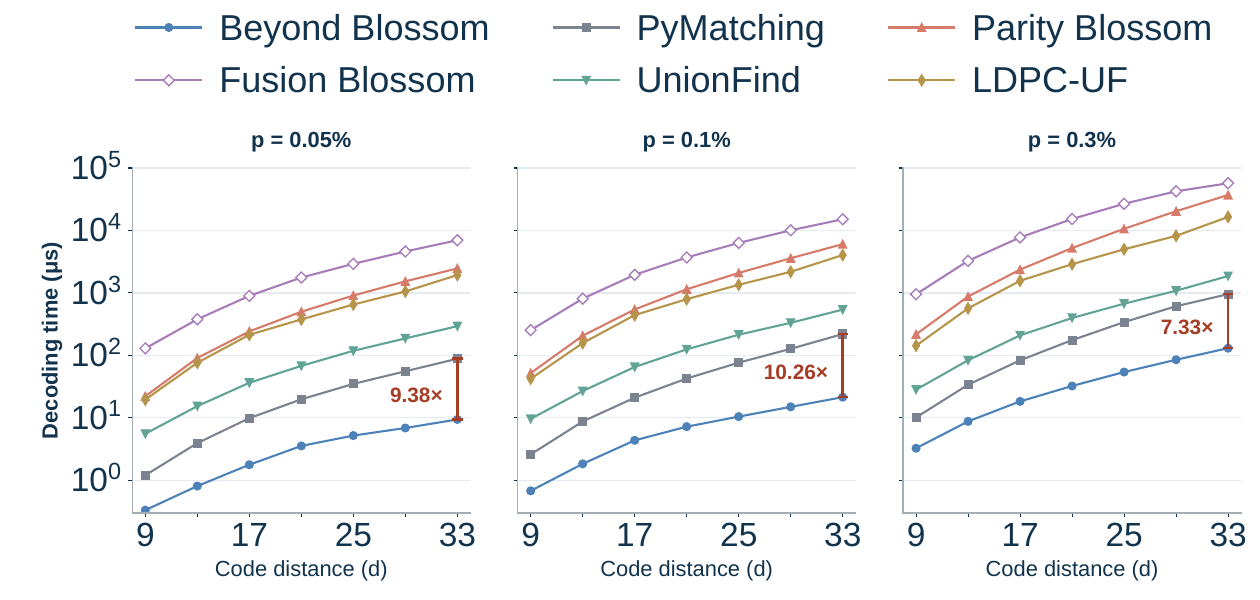}
  \caption{Amortized decoding time versus code distance for the compared decoders at three physical error rates, with \(d\) measurement rounds. Filled markers denote \(10^6\) shots. Open markers denote \(10^4\) shots, used when this setting would require prohibitively long runtimes.}
  \label{fig:decoder-comparison}
\end{figure}

Figure~\ref{fig:decoder-comparison} shows that Beyond Blossom achieves the lowest amortized decoding time among the evaluated implementations across all 21 settings.
Its largest speedups over PyMatching are \(\PerfValue{xeon.decoder-comparison.p0p05.best.vs-pymatching.max-speedup}\times\), \(\PerfValue{xeon.decoder-comparison.p0p1.best.vs-pymatching.max-speedup}\times\), and \(\PerfValue{xeon.decoder-comparison.p0p3.best.vs-pymatching.max-speedup}\times\) at the three respective error rates.
Even the Union-Find implementations, which typically trade decoding accuracy for simpler computation, are slower in these settings.
Beyond Blossom thus achieves the best performance among the evaluated CPU implementations while preserving the exact MWPM objective.

The Beyond Blossom curve uses the most suitable configuration for each setting, employing one to eight physical cores.
Small workloads typically favor single-threaded predecoding, whereas larger blocks benefit further from the \emph{Pipeline-Wide CPU Parallelism for Short Decoding Blocks} described in Section~\ref{sec:multilevel-parallelism}.
The following ablation explains this transition.

\subsection{Incremental Ablation}
\label{sec:eval-ablation}

\begin{figure*}[!tbp]
  \centering
  \makebox[\textwidth][c]{%
        \includegraphics[width=1\textwidth]{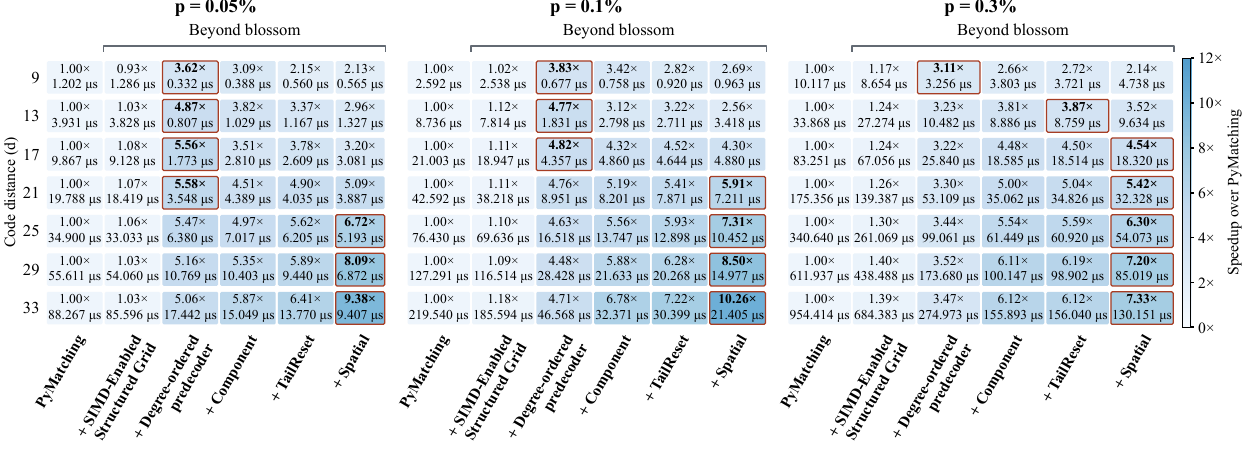}%
    }
  \caption{Incremental ablation of Beyond Blossom across code distances and physical error rates. Each cell reports amortized decoding time and speedup relative to PyMatching. PyMatching, the structured SIMD backend, and predecoding use one physical core. The successive parallel configurations use two, four, and eight physical cores.}
  \label{fig:incremental-ablation}
\end{figure*}

Figure~\ref{fig:incremental-ablation} evaluates our optimizations incrementally against the state-of-the-art single-threaded PyMatching baseline.
We first introduce the structured grid and SIMD computation (Section~\ref{sec:structured-grid-simd}), followed by degree-ordered predecoding (Section~\ref{sec:predecoding}), while retaining single-core execution.
We then enable component parallelism, tail-processing optimizations with overlapped reset, and intra-component predecoding parallelism (Section~\ref{sec:multilevel-parallelism}), using two, four, and eight physical cores, respectively.
All speedups shown in Fig.~\ref{fig:incremental-ablation} are measured relative to PyMatching; differences between adjacent configurations show the incremental benefit of each optimization.

For this ablation, all six variants process the same \(10^6\) independent shots per setting. We report amortized decoding time per shot after warm-up.

\paragraph{Architecture-aware optimizations}
The structured representation and SIMD dual-phase computation provide larger benefits as the workload increases. Fixed-offset access to compact state reduces pointer chasing and improves cache-line utilization during neighborhood scans. 
At \(d=33,p=0.3\%\), the corresponding L2/L3 miss rates decreased from \(\PerfValue{xeon.ablation.d33.p0p3.pymatching.L2miss}\%/\PerfValue{xeon.ablation.d33.p0p3.pymatching.L3miss}\%\) for PyMatching to \(\PerfValue{xeon.ablation.d33.p0p3.structured-simd.L2miss}\%/\PerfValue{xeon.ablation.d33.p0p3.structured-simd.L3miss}\%\) for Beyond Blossom.
At the same setting, decoding time decreases from \(\PerfValue{xeon.ablation.d33.p0p3.pymatching.mean-time-us}\,\mu\mathrm{s}\) to \(\PerfValue{xeon.ablation.d33.p0p3.structured-simd.mean-time-us}\,\mu\mathrm{s}\), yielding a \(\PerfValue{xeon.ablation.d33.p0p3.structured-simd.vs-pymatching.speedup}\times\) speedup.

At \(d=9,p=0.05\%\), decoding time instead increases slightly, from \(\PerfValue{xeon.ablation.d9.p0p05.pymatching.mean-time-us}\,\mu\mathrm{s}\) to \(\PerfValue{xeon.ablation.d9.p0p05.structured-simd.mean-time-us}\,\mu\mathrm{s}\).
The structured SIMD backend introduces input-organization, vector-packing, masking, and minimum-reduction costs.
Vectorized scans also evaluate some candidates that scalar branches can discard early.
For very small workloads, these costs can outweigh the benefits of regular memory access and parallel arithmetic.

\paragraph{Degree-ordered predecoding}
Adding predecoding substantially reduces decoding time across all 21 settings, providing approximately \(\PerfValue{xeon.ablation.predecoding.vs-structured-simd.min-speedup}\text{--}\PerfValue{xeon.ablation.predecoding.vs-structured-simd.max-speedup}\times\) additional speedup over the preceding configuration.
Resolving defects locally reduces the work entering the full MWPM search.
This optimization is particularly effective at low error rates, where local error patterns are prevalent.

\paragraph{Pipeline-wide CPU parallelism}
Independent \(X/Z\) components provide an opportunity for parallel execution, but exploiting it requires enough work to offset coordination costs.
At \(p=0.1\%\), component parallelism increases decoding time from \(\PerfValue{xeon.ablation.d13.p0p1.predecoding.mean-time-us}\,\mu\mathrm{s}\) to \(\PerfValue{xeon.ablation.d13.p0p1.component-parallel.mean-time-us}\,\mu\mathrm{s}\) for \(d=13\), but reduces it from \(\PerfValue{xeon.ablation.d33.p0p1.predecoding.mean-time-us}\,\mu\mathrm{s}\) to \(\PerfValue{xeon.ablation.d33.p0p1.component-parallel.mean-time-us}\,\mu\mathrm{s}\) for \(d=33\), yielding a \(\PerfValue{xeon.ablation.d33.p0p1.component-parallel.vs-predecoding.speedup}\times\) speedup.
As the error rate increases, component parallelism becomes beneficial at smaller code distances.

This crossover reflects the per-input costs of task dispatch, synchronization, and result merging.
Small decoding blocks already have short sequential execution times, making these costs difficult to amortize.
Serial input organization and unequal \(X/Z\) workloads further limit scaling: synchronization waits for the slower component, so practical speedups generally remain below \(2\times\).

Tail-processing optimizations and reset overlap further reduce decoding time in \PerfValue{xeon.ablation.tail-reset.num-improved-settings} settings.
As decoding approaches microsecond timescales, state maintenance accounts for an appreciable fraction of execution time.
Overlapping cleanup of the previous input with decoding of the next reduces its exposed cost.

Finally, intra-component predecoding parallelism provides additional gains for larger workloads.
At \(d=33,p=0.3\%\), decoding time decreases from \(\PerfValue{xeon.ablation.d33.p0p3.tail-reset.mean-time-us}\,\mu\mathrm{s}\) to \(\PerfValue{xeon.ablation.d33.p0p3.parallel-predecoding.mean-time-us}\,\mu\mathrm{s}\), reaching a cumulative \(\PerfValue{xeon.ablation.d33.p0p3.parallel-predecoding.vs-pymatching.speedup}\times\) speedup over PyMatching.
Smaller workloads can regress because of partitioning, synchronization, and sequential conflict-band processing.

Among the evaluated open-source CPU decoders, Beyond Blossom is the only implementation whose parallel execution surpasses the state-of-the-art single-threaded PyMatching baseline on short \(d\)-round decoding blocks.
Its additional gains over our optimized sequential configuration demonstrate the effectiveness of directing parallelism toward local predecoding and pipeline operations.

\subsection{Understanding the Speedup}
\label{sec:eval-predecoding}

To isolate the effect of predecoding, we compare no predecoding, simple predecoding, and degree-ordered predecoding in Beyond Blossom, with all other settings fixed.
The simple baseline omits degree ordering and uses equal-rate growth, with each endpoint contributing half the edge weight to a direct match.
Our method orders defects by candidate degree and allocates growth according to the current dual constraints (Section~\ref{sec:predecoding}).

Table~\ref{tab:predecoder-comparison} reports complete decoding time, the mean number of defects left unmatched after predecoding, and the fraction of defects matched during predecoding. At \(p=0.3\%\), degree-ordered predecoding matches approximately \(\PerfValue{xeon.predecoder-comparison.p0p3.degree-ordered.matched-percent.min}\%\)--\(\PerfValue{xeon.predecoder-comparison.p0p3.degree-ordered.matched-percent.max}\%\) of the initial defects. For \(d=33\), the mean number of unmatched defects decreases from \(\PerfValue{xeon.predecoder-comparison.d33.p0p3.none.mean-remaining-syndromes}\) to \(\PerfValue{xeon.predecoder-comparison.d33.p0p3.degree-ordered.mean-remaining-syndromes}\), substantially reducing the number of unmatched roots entering the full solver. Complete decoding time, including predecoding when enabled, decreases from \(\PerfValue{xeon.predecoder-comparison.d33.p0p3.none.mean-time-us}\,\mu\mathrm{s}\) without predecoding to \(\PerfValue{xeon.predecoder-comparison.d33.p0p3.degree-ordered.mean-time-us}\,\mu\mathrm{s}\) with degree-ordered predecoding.

\begin{table}[!htbp]
  \centering
  \caption{Serial predecoding at \(p=0.3\%\) with \(d\) measurement rounds}
  \label{tab:predecoder-comparison}
  \footnotesize
  \setlength{\tabcolsep}{4pt}
  \renewcommand{\arraystretch}{1.12}
  \begin{tabular*}{\columnwidth}{@{\extracolsep{\fill}}rrrr@{}}
    \toprule
    \multicolumn{4}{l}{\textit{(a) Amortized decoding time (\(\mu\mathrm{s}/\mathrm{shot}\))}} \\
    \addlinespace
    \(d\) & None & Simple & \textbf{Degree-ordered} \\
    \midrule
    9
      & \PerfValue{xeon.predecoder-comparison.d9.p0p3.none.mean-time-us}
      & \PerfValue{xeon.predecoder-comparison.d9.p0p3.coordinate.mean-time-us}
      & \textbf{\PerfValue{xeon.predecoder-comparison.d9.p0p3.degree-ordered.mean-time-us}} \\
    13
      & \PerfValue{xeon.predecoder-comparison.d13.p0p3.none.mean-time-us}
      & \PerfValue{xeon.predecoder-comparison.d13.p0p3.coordinate.mean-time-us}
      & \textbf{\PerfValue{xeon.predecoder-comparison.d13.p0p3.degree-ordered.mean-time-us}} \\
    17
      & \PerfValue{xeon.predecoder-comparison.d17.p0p3.none.mean-time-us}
      & \PerfValue{xeon.predecoder-comparison.d17.p0p3.coordinate.mean-time-us}
      & \textbf{\PerfValue{xeon.predecoder-comparison.d17.p0p3.degree-ordered.mean-time-us}} \\
    21
      & \PerfValue{xeon.predecoder-comparison.d21.p0p3.none.mean-time-us}
      & \PerfValue{xeon.predecoder-comparison.d21.p0p3.coordinate.mean-time-us}
      & \textbf{\PerfValue{xeon.predecoder-comparison.d21.p0p3.degree-ordered.mean-time-us}} \\
    25
      & \PerfValue{xeon.predecoder-comparison.d25.p0p3.none.mean-time-us}
      & \PerfValue{xeon.predecoder-comparison.d25.p0p3.coordinate.mean-time-us}
      & \textbf{\PerfValue{xeon.predecoder-comparison.d25.p0p3.degree-ordered.mean-time-us}} \\
    29
      & \PerfValue{xeon.predecoder-comparison.d29.p0p3.none.mean-time-us}
      & \PerfValue{xeon.predecoder-comparison.d29.p0p3.coordinate.mean-time-us}
      & \textbf{\PerfValue{xeon.predecoder-comparison.d29.p0p3.degree-ordered.mean-time-us}} \\
    33
      & \PerfValue{xeon.predecoder-comparison.d33.p0p3.none.mean-time-us}
      & \PerfValue{xeon.predecoder-comparison.d33.p0p3.coordinate.mean-time-us}
      & \textbf{\PerfValue{xeon.predecoder-comparison.d33.p0p3.degree-ordered.mean-time-us}} \\
    \addlinespace
    \multicolumn{4}{l}{\textit{(b) Mean unmatched defects/shot (matched \%)}} \\
    \addlinespace
    \(d\) & None & Simple & \textbf{Degree-ordered} \\
    \midrule
    9
      & \PerfValue{xeon.predecoder-comparison.d9.p0p3.none.mean-remaining-syndromes}
      & \PerfValue{xeon.predecoder-comparison.d9.p0p3.coordinate.mean-remaining-syndromes}
        (\PerfValue{xeon.predecoder-comparison.d9.p0p3.coordinate.matched-percent}\%)
      & \textbf{\PerfValue{xeon.predecoder-comparison.d9.p0p3.degree-ordered.mean-remaining-syndromes}}
        (\PerfValue{xeon.predecoder-comparison.d9.p0p3.degree-ordered.matched-percent}\%) \\
    13
      & \PerfValue{xeon.predecoder-comparison.d13.p0p3.none.mean-remaining-syndromes}
      & \PerfValue{xeon.predecoder-comparison.d13.p0p3.coordinate.mean-remaining-syndromes}
        (\PerfValue{xeon.predecoder-comparison.d13.p0p3.coordinate.matched-percent}\%)
      & \textbf{\PerfValue{xeon.predecoder-comparison.d13.p0p3.degree-ordered.mean-remaining-syndromes}}
        (\PerfValue{xeon.predecoder-comparison.d13.p0p3.degree-ordered.matched-percent}\%) \\
    17
      & \PerfValue{xeon.predecoder-comparison.d17.p0p3.none.mean-remaining-syndromes}
      & \PerfValue{xeon.predecoder-comparison.d17.p0p3.coordinate.mean-remaining-syndromes}
        (\PerfValue{xeon.predecoder-comparison.d17.p0p3.coordinate.matched-percent}\%)
      & \textbf{\PerfValue{xeon.predecoder-comparison.d17.p0p3.degree-ordered.mean-remaining-syndromes}}
        (\PerfValue{xeon.predecoder-comparison.d17.p0p3.degree-ordered.matched-percent}\%) \\
    21
      & \PerfValue{xeon.predecoder-comparison.d21.p0p3.none.mean-remaining-syndromes}
      & \PerfValue{xeon.predecoder-comparison.d21.p0p3.coordinate.mean-remaining-syndromes}
        (\PerfValue{xeon.predecoder-comparison.d21.p0p3.coordinate.matched-percent}\%)
      & \textbf{\PerfValue{xeon.predecoder-comparison.d21.p0p3.degree-ordered.mean-remaining-syndromes}}
        (\PerfValue{xeon.predecoder-comparison.d21.p0p3.degree-ordered.matched-percent}\%) \\
    25
      & \PerfValue{xeon.predecoder-comparison.d25.p0p3.none.mean-remaining-syndromes}
      & \PerfValue{xeon.predecoder-comparison.d25.p0p3.coordinate.mean-remaining-syndromes}
        (\PerfValue{xeon.predecoder-comparison.d25.p0p3.coordinate.matched-percent}\%)
      & \textbf{\PerfValue{xeon.predecoder-comparison.d25.p0p3.degree-ordered.mean-remaining-syndromes}}
        (\PerfValue{xeon.predecoder-comparison.d25.p0p3.degree-ordered.matched-percent}\%) \\
    29
      & \PerfValue{xeon.predecoder-comparison.d29.p0p3.none.mean-remaining-syndromes}
      & \PerfValue{xeon.predecoder-comparison.d29.p0p3.coordinate.mean-remaining-syndromes}
        (\PerfValue{xeon.predecoder-comparison.d29.p0p3.coordinate.matched-percent}\%)
      & \textbf{\PerfValue{xeon.predecoder-comparison.d29.p0p3.degree-ordered.mean-remaining-syndromes}}
        (\PerfValue{xeon.predecoder-comparison.d29.p0p3.degree-ordered.matched-percent}\%) \\
    33
      & \PerfValue{xeon.predecoder-comparison.d33.p0p3.none.mean-remaining-syndromes}
      & \PerfValue{xeon.predecoder-comparison.d33.p0p3.coordinate.mean-remaining-syndromes}
        (\PerfValue{xeon.predecoder-comparison.d33.p0p3.coordinate.matched-percent}\%)
      & \textbf{\PerfValue{xeon.predecoder-comparison.d33.p0p3.degree-ordered.mean-remaining-syndromes}}
        (\PerfValue{xeon.predecoder-comparison.d33.p0p3.degree-ordered.matched-percent}\%) \\
    \bottomrule
  \end{tabular*}
\end{table}

\subsection{Arm Portability and Processing Throughput}
\label{sec:eval-arm}

We evaluate Beyond Blossom on an Apple M4 Pro using the same workloads, selecting a suitable parallelization strategy for each combination of code distance \(d\) and physical error rate \(p\).
Each setting uses \(10^6\) saved independent inputs; reported times are the median of three complete-batch mean times per shot.
Beyond Blossom accelerates decoding across all 21 settings, with maximum speedups over PyMatching of \(6.36\times\), \(6.41\times\), and \(5.02\times\) at \(p=0.05\%\), \(0.1\%\), and \(0.3\%\), respectively.

\begin{figure}[!htbp]
  \centering
  \includegraphics[width=\linewidth]{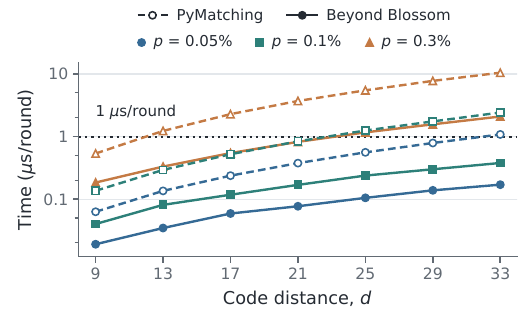}
  \caption{Decoding time per measurement round on Apple M4 Pro (logarithmic y-axis).
    The horizontal line marks \(1\,\mu\mathrm{s}/\mathrm{round}\).}
  \label{fig:m4-per-round}
\end{figure}

To assess processing capacity, we divide the reported block decoding time by its \(d\) measurement rounds.
Figure~\ref{fig:m4-per-round} compares the resulting time per round with a \(1\,\mu\mathrm{s}\) measurement cycle, representative of the interval between measurement rounds in superconducting quantum processors~\cite{google2025belowthreshold}.
Beyond Blossom meets this processing budget for every tested distance from \(d=9\) to \(d=33\) at both \(p=0.05\%\) and \(p=0.1\%\).
PyMatching reaches \(d=29\) and \(d=21\), respectively.
At \(p=0.3\%\), Beyond Blossom extends the supported range to \(d=21\), compared with \(d=9\) for PyMatching.
Overall, Beyond Blossom meets the budget in 18 of the 21 settings, compared with 11 for PyMatching.

On the M4 Pro, Beyond Blossom achieves lower amortized decoding time than on Xeon.
The faster platform therefore allows the decoder to keep pace with the \(1\,\mu\mathrm{s}\) measurement cycle over a wider range of workloads.

\FloatBarrier

\section{Conclusion}
\label{sec:conclusion}

We presented Beyond Blossom, a CPU-based MWPM decoder for surface codes.
It combines degree-ordered local predecoding with structured graph storage,
SIMD computation, and parallel execution to reduce decoding time while
preserving MWPM optimality. On an x86 CPU, it achieves a \(\PerfValue{legacy.single-core.vs-pymatching.geomean-speedup}\times\)
geometric-mean speedup on a single core and up to a  \(\PerfValue{xeon.decoder-comparison.p0p1.best.vs-pymatching.max-speedup}\times\) speedup
with eight cores over single-threaded PyMatching. Results on Arm also
demonstrate portability. Future work will focus on streaming decoding,
GPU acceleration for batched workloads, and integration with quantum
control systems to evaluate end-to-end feedback latency.

\bibliography{references}
\end{document}